\documentclass{article}
\usepackage{spconf,amsmath,graphicx,diagbox,amssymb,algorithm,algpseudocode,amsmath,amssymb,mathtools,nccmath}

\def\x{{\mathbf x}}
\def\hatx{{\mathbf{\hat{x}}}}
\def\v{{\mathbf v}}
\def\s{{\mathbf s}}
\def\b{{\mathbf b}}
\def\a{{\mathbf a}}

\def\X{{\mathbf X}}
\def\hatX{{\mathbf{\hat{X}}}}
\def\V{\mathbf V}

\def\R{{\mathbb R}}
\def\B{{\mathbb B}}
\def\rhohat{\hat{\rho}}

\title{SECP: A SPEECH ENHANCEMENT-BASED CURATION PIPELINE FOR SCALABLE ACQUISITION OF CLEAN SPEECH}

\name{Adam Sabra, Cyprian Wronka, Michelle Mao, Samer Hijazi}
\address{Cisco Systems, Inc. \\
\texttt{\{asabra, cwronka, maoxh, hijazy\}@cisco.com}
}
\begin{document}
%

\maketitle

\begin{abstract}
As more speech technologies rely on a supervised deep learning approach with clean speech as the ground truth, a methodology to onboard said speech at scale is needed. However, this approach needs to minimize the dependency on human listening and annotation, only requiring a human-in-the-loop when needed. In this paper, we address this issue by outlining Speech Enhancement-based Curation Pipeline (SECP) which serves as a framework to onboard clean speech. This clean speech can then train a speech enhancement model, which can further refine the original dataset and thus close the iterative loop. By running two iterative rounds, we observe that enhanced output used as ground truth does not degrade model performance according to $\Delta_{PESQ}$, a metric used in this paper. We also show through comparative mean opinion score (CMOS) based subjective tests that the highest and lowest bound of refined data is perceptually better than the original data.
\end{abstract}


\section{Introduction}
\label{sec:intro}

As speech-based technologies rely on deep learning, the need for clean speech rises proportionally. Onboarding clean speech data is a non-trivial process which generally requires many manual hours to listen and annotate clean speech. However, this proves to be both an expensive and intensive process. Having an automated process is needed to address both scaling up speech acquisition while simultaneously allowing manual annotation to have an outsized impact.

There are annotation processes in other modalities such as computer vision which enable annotation on the pixel basis \cite{RL_Semantic_Segmentation}. Furthermore, speech recognition tasks (which partially exists in the audio modality) allow for direct editing of low-performing transcribed speech which can then be used to retrain the model \cite{AL_ASR, AL_ASR_Unsupervised}. Despite both annotation methods working well in their respective domain, there is very little overlap in what can be done to “listen” and locate clean speech, which is imperative for audio-based supervised learning tasks like speech enhancement (SE) and text-to-speech (TTS).

\begin{figure}[t]
    \centering
    \centerline{\includegraphics[width=6.5cm]{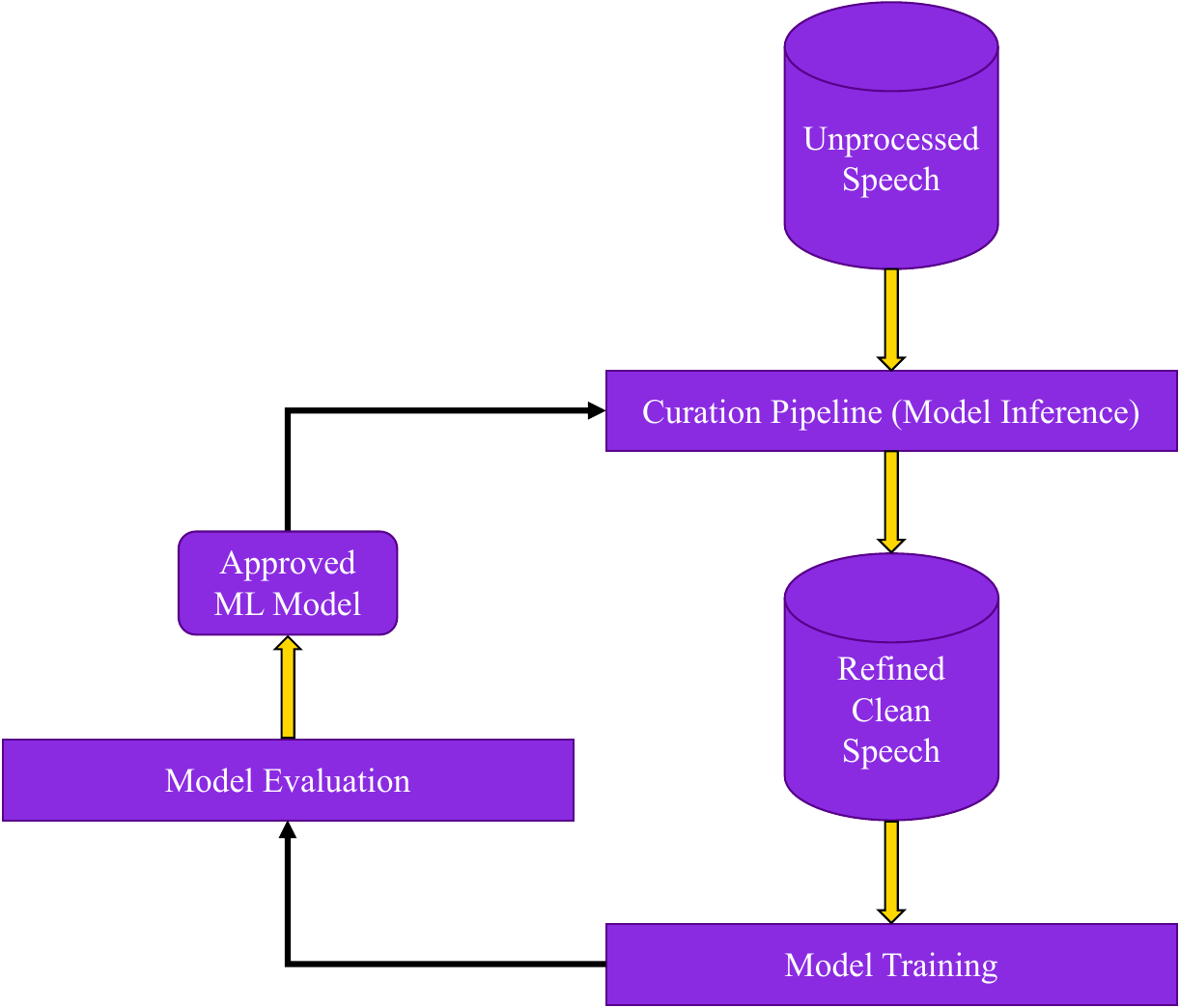}}
    \caption{A high level overview of the proposed iterative process in which both the data and speech enhancement model improve each other.}
    \label{fig:overview_pipeline}
\end{figure}

In this paper, we aim to outline SECP – the Speech Enhancement-based Curation Pipeline. The pipeline has two main objectives: cleaning speech in large sets of audio files reliably, and formatting said speech for further downstream use cases. The goal of these two objectives is to develop a process in which the acquisition of clean speech and the training of a speech enhancement model improve each other symbiotically and iteratively.

SE and TTS architectures described in research papers are generally trained in a supervised fashion with the ground truth being anechoic clean speech. These tasks often utilize datasets such as LJSpeech \cite{ljspeech} and Voice Bank \cite{VoiceBank}. While these datasets serve as a useful benchmark, they are not sufficient for generalization.

For the development of data-driven models tailored to speech-oriented applications, it is ideal to construct a substantial dataset consisting of a large speech corpus to mitigate potential bias. A lack of generalization, in these cases, can be looked at as favoring one language over another, removing speech or noise when not necessary, as well as discriminating against speech across factors such as gender, age, accent, and emotions.

\begin{algorithm*}
\caption{Overview of the SECP Algorithm}

\begin{algorithmic}
\Require $f_s$ \Comment{Specify sampling rate}
\State $N$ \Comment{Duration of saved curated sample (in seconds)}
\State $w_l$ \Comment{Duration of frame size (in seconds)}
\State $s_{th}$ \Comment{$\hat{\rho}$ threshold to filter out high noise frames}
\State $b_w$ \Comment{Frequency bandwidth threshold to filter out low-bandwidth frames}
\State $\texttt{SE}$ \Comment{Speech Enhancement Model}
\State $\texttt{VAD}$ \Comment{Voice Activity Detection Model}
\Ensure $\x \in \mathbb{R}^{1 \times (f_s \cdot L)}$
\end{algorithmic}

\begin{algorithmic}[1]
\For{number of files in dataset}
\vspace{0.5mm}
\State $\hatx \gets \texttt{SE}(\x)$, where $\hatx \in \mathbb{R}^{1 \times (f_s \cdot L)}$

\vspace{0.5mm}
\State $\v \gets \texttt{VAD}(\hatx)$, where $\v \in \mathbb{B}^{1 \times (f_s \cdot L)}$ and $\mathbb{B} \coloneqq \{0,1\}$

\vspace{0.5mm}
\State $\X, \hatX, \V \gets \text{Reshape } \x, \hatx, \v$, where $\{ \X, \hatX, \V \} \in \mathbb{R}^{ (f_s \cdot w_l) \times \frac{L}{w_l} }$

\vspace{0.5mm}
\State $\tilde{\rho} \gets \hat{\rho}(\X_{*,l}, \hatX_{*,l}, \V_{*,l})$, for all $ l = 1, \dots, \frac{L}{w_l}$ and $\tilde{\rho} \in \mathbb{R}^{1 \times \frac{L}{w_l}}$

\State $\s \gets \hat{\rho} > s_{th}, \text{where } \s \in \mathbb{B}^{1 \times \frac{L}{w_l}}$

\State $\b \gets f_c \geq b_w$ for all frames in $\hatX$, where $\b \in \mathbb{B}^{1 \times \frac{L}{w_l}}$.

\State $\a \gets \s \dot\wedge \b$, where $\a \in \mathbb{B}^{1 \times \frac{L}{w_l}}$

\State Locate all $N$ seconds of contiguous, non-overlapping, and approved frames in $\a$.

\EndFor
\end{algorithmic}
\end{algorithm*}

The main objective of this pipeline is to obtain more high-quality clean speech which allows us to improve the quality and distribution of our speech data. The improvement in our datasets can then be used to train a new model, thus reprocessing the dataset once again and onboard more clean data. As opposed to a scalable data augmentation engine with a fixed speech corpus proposed by \cite{SND}, we propose a methodology aimed to strictly increase the clean speech corpus available for research teams to utilize for SE and TTS research tasks.

\section{Methodology}
\label{sec:methodology}

\subsection{Neural Network Models}

In the SECP, we leverage speech enhancement and voice activity detection (VAD) neural networks to ensure our processing is done on clean speech. In this subsection, we aim to outline the design of these two networks and the function they serve in the pipeline.

Our speech enhancement model is a U-Net based Convolutional Neural Network \cite{UNet_Image_Segmentation, Causal_SE, Fully_CNN_SE, P_V__2019}, operating on sampling rate $f_s$. The network learns to produce a magnitude mask in the time-frequency domain and we found a non-causal network to serve us best for the objective of curation. 

We computed the STFT, and subsequent ISTFT, using window size $w_s$ with a hop size of $0.25w_s$. The mask generated takes in context from both past and future frames to make a prediction on the center frame. Although the model is non-causal, we developed it such that it is able to process files that are tens of minutes and even over an hour long. This allows us to run enhancement in an offline process without worrying about real-time constraints. This network is the model used for experimentation in this paper.

Furthermore, we develop a convolutional-based non-causal VAD neural network that is trained on noisy speech \cite{noncausalVad, noisyVad}. Our VAD takes in both left and right context to predict the speech presence probability (SPP) on window length $w_v$ as shown in \cite{causalVAD}. After determining the binary decision of speech presence in the window, we extrapolate the decision to all samples in the window before continuing on to the next one. By doing this, we ensure the VAD decision is shape aligned with the input and enhanced audio. The VAD is not retrained throughout rounds and is only used for inference.

\begin{figure*}[h]
    \centering
    \centerline{\includegraphics[width=17cm]{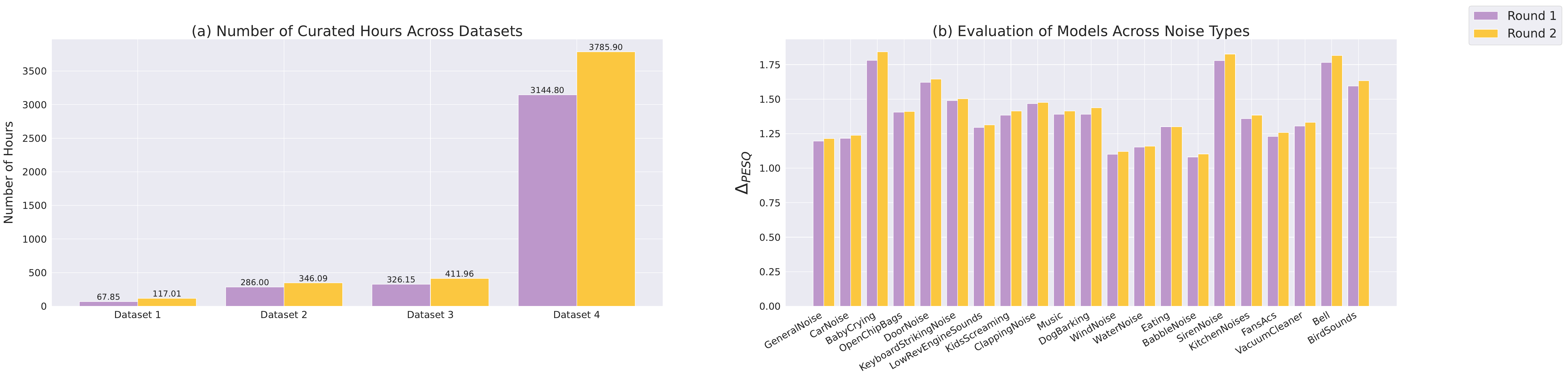}}
    \caption{The number of accepted curated hours between rounds (a) with the comparison of  $\Delta_{PESQ}$ scores between model training rounds across various noise types (b).}
    \label{fig:spenh_and_hours}
\end{figure*}

\subsection{SECP Algorithm}

When designing the pipeline, we needed to ensure that the criteria of the formatted speech was aligned with the criteria of publicly available TTS \cite{Hifi_TTS, LibriTTS} and SE datasets \cite{DNS} as the ground truth outputs for both problems require clean speech. Upon investigating datasets for both problems, full band data with a high SNR served as a minimum criterion for acceptance.

SECP leverages two machine learning models to curate data, a VAD model and  SE model, both outlined in Section 2.1. Furthermore, the pipeline requires design parameters for formatting the curated data. We aim to utilize these models as modular components to the pipeline to ensure flexibility.

The algorithm takes in audio file $\x$, where $\x \in \R^{1 \times (f_s \cdot L)}$ where $L$ is the duration of the audio file in seconds. We first enhance the audio to produce $\hatx$ and run the VAD on the enhanced file to get the sample-wise binary decisions, $\v$, where $\hatx \in \R^{1 \times (f_s \cdot L)}$ and $\v \in \B^{1 \times (f_s \cdot L)}$. We define $\B$ as the set of binary numbers, or $\B \coloneqq \{0, 1\}$.

Once we have $\x, \hatx, \text{and } \v$, we then reshape the vectors into frames of length $w_l \cdot f_s$, where $w_l$ is a parameter for specifying the frame length. In our experiments, $w_l$ is chosen to be 1 second long. $\x, \hatx, \text{and } \v$ then become $\X, \hatX, \text{and } \V$, respectively where $\{ \X, \hatX \} \in \R^{(w_l \cdot f_s) \times \frac{L}{w_l}}$ and $\V \in \B^{(w_l \cdot f_s) \times \frac{L}{w_l}}$. 

By framing our vectors, we are able to perform our $SNR$ calculation on each window, narrowing the search for high $SNR$ speech. To find high $SNR$ speech segments, we formulate $\rhohat$ as an estimate of $SNR$. We calculate and define $\hat{\rho}$ as the following:

\begin{multline}
\label{eq:snr_l}
\hat{\rho}_l(\x_l, \hatx_l, \v_l) = \\
\begin{cases}
RMSdB(\hatx_l) - RMSdB(\x_l - \hatx_l), & \text{if $\mathbb{E}(\v_l) \geq 0.5$} \\
- \infty, & \text{otherwise}
\end{cases}
\end{multline}

\begin{equation}
    RMSdB(\x) = 20 \log_{10} \left( \sqrt{\frac{1}{n} \sum_{i=0}^{n}{x_{i}^2}} \text{ } \right)
\end{equation}

Where $\hatx_l$ represents the frame of enhanced speech, $\x_l$ represents the frame of the corresponding unprocessed speech, and $\v_l$ corresponds to the sample-wise VAD decision of the entire frame, where $ \{ \x_l, \hatx_l, \v_l \} \in \mathbb{R}^{1 \times \frac{L}{w_l}}$. Simply we perform the $SNR$ estimation using each column of $\X, \hatX,$ and $\V$. 

The core assumption in our formulation is that the speech enhancement model will remove noise and preserve speech, and thus the difference will only contain speech. Because $\rhohat$ serves as an estimate to the actual $SNR$, investigating its performance was necessary for robustness. To do this, we tested it on various situations to ensure that we would not get undesirable segments labelled as clean speech. 

We ran internal listening and annotation to determine that an $SNR$ threshold, notated as $s_{th}$ in Algorithm 1, of 20 dB was a sufficient threshold to use to locate clean speech with adequate noise removal. More specifically, we found that when $\rhohat \geq 20$, the enhanced speech files removed noise to a degree that not only was unperceivable by human annotators, but also was not visible in spectrogram visualizations. For our experiments, we find and calculate $\s$, a vector of all binary decisions of segments containing $\rhohat \geq 20$ dB, where $\s \in \mathbb{B}^{1 \times \frac{L}{w_l}}$.

To detect and filter out low-bandwidth content, we leverage simple signal processing techniques to analyze binned spectrograms to allow for time-windowed analysis of the content. The cut-off frequency, $f_c$, per each time window is calculated and used to determine the bandwidth of the frame. If $f_c \geq b_w$, where $b_w$ is the specified bandwidth, the frame is preserved. This is used to filter out low-bandwidth content and preserve high-bandwidth content. We assign all binary results of bandwidth detection to $\b$, where $\b \in \mathbb{B}^{1 \times \frac{L}{w_l}}.$

After calculating vectors $\s$ and $\b$, we use an entry wise logical AND operation, notated as $\dot\wedge$. We assign the entry-wise logical AND output as $\a$, which determines all of our approved, high-quality segments in our input audio file. Every non-overlapping $N$ seconds of contiguous accepted segments would then be considered as one curated sample. In our experiments, $N$ is set to 12 seconds. All the curated sample’s relevant metadata will be saved, and this process is repeated until the entire dataset is fully curated.

\begin{figure*}[ht!]
    \centering
    \centerline{\includegraphics[width=18cm]{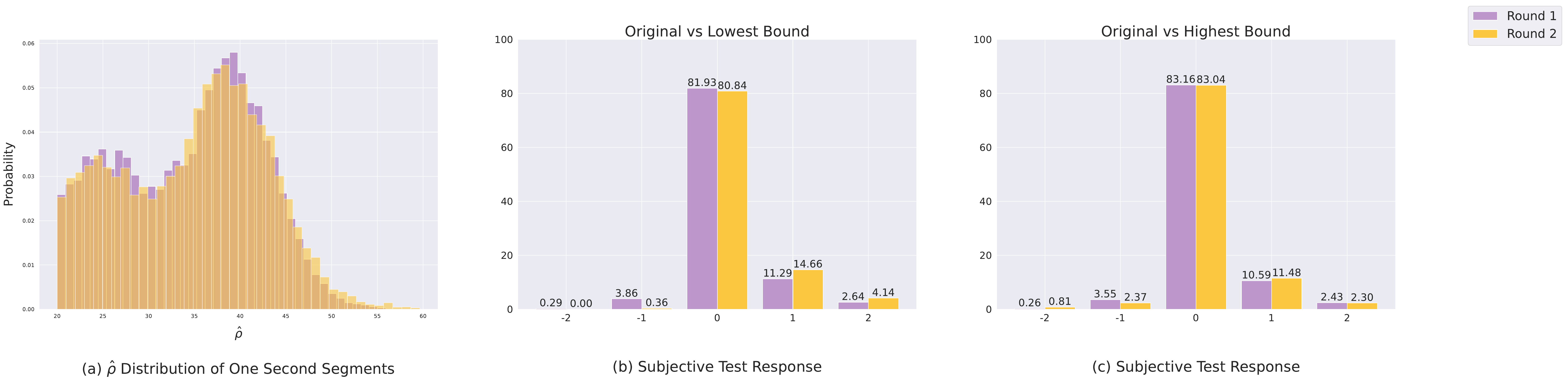}}
    \caption{$\hat{\rho}$ distribution of approved one second segements between rounds (a), as well as subjective test results comparing lowest bound of performance (b) and the highest bound of performance (c) to the original unprocessed files.}
    \label{fig:spenh_and_hours}
\end{figure*}

\subsection{Iterative Training and Data Onboarding}

The methodology of implementing SECP aims to be a systematic one. This means that even if a research team does not have a large corpus of data, starting with a small seed round to train the first Speech Enhancement model is sufficient. To show this, we first began by training our model on a small set of data that we annotated as clean. This model then processes four internal dataset fed into SECP. This newly curated data serves as ground truth for the training of a second model which then formats the same unprocessed dataset and repeats the workflow all over again.

\section{Results}
\label{sec:results}

We use our non-causal wideband model outlined in Section 3.1 for two rounds of curation and training with no architectural changes. We aim to evaluate our model objectively, for the direct comparison between models in denoising performance, and subjectively, to ensure that objective measurement is correlated with human perception.

\subsection{Objective Model Performance}

Upon investigating the curation performance outlined in Figure 2a, we see an improvement across all four internal sets. Aside from the 72.4\% increase in Dataset 1, the other three sets increased by 21.01\%, 26.30\%, and 20.03\%, respectively.

Upon analysis of our denoising model, we studied its performance in removing noise from speech across 21 noise categories shown in Figure 2b. Each noise set contains 1000 clean speech files with noise injected in them. The degree of noise that is added to the speech file is determined by a Rayleigh distribution. The test sets in 2b are fixed between rounds. 

To measure the models’ performances in denoising, we define $\Delta_{PESQ}$ as the following:

\begin{equation}
    \Delta_{PESQ} = PESQ(\hatx, \x) - PESQ( \Tilde{\x}, \x)
\end{equation}

Where $\Tilde{\x}$ represents the noisy speech, $\x$ represents the clean speech, $\hatx$ represents the enhanced noisy speech files, and $PESQ$ is defined in \cite{PESQ}. The higher the $\Delta_{PESQ}$ performance, the larger the difference between the noisy and enhanced files are. In Figure 2b, we see a general improvement in $\Delta_{PESQ}$ between rounds.

\subsection{Subjective Model Performance}

When observing the $\hat{\rho}$ distribution in Figure 3 (a), we notice a general decrease in $SNR_c \leq 30$ dB as well as an increase in $SNR_c \geq 45$ dB. This behavior is ideal as we aim to use the pipeline to shift the $\hat{\rho}$ distribution towards cleaner speech, or higher $\rhohat$ estimations. 

However, to show that the data is of equal, or better, quality at these lower and upper bounds, we utilize subjective testing to verify a perceptual difference between processing rounds. To do this, we leveraged $\hat{\rho}$ and time stamp metadata saved from the pipeline to see if the how the lowest ($SNR_c \leq 30$ dB) and highest ($SNR_c \geq 45$ dB) bounds of both rounds compare to the original unprocessed files. 

We can filter curated files by $\rhohat$ estimations. Because each curated file was set to $N = 12$ and $w_l = 1$, each curated file has a $\rhohat$ array of length 12. We can filter on the lowest and highest $\rhohat$ segments to locate files that satisfy this criteria.
Furthermore, by leveraging the timestamp metadata, we are able to snip out the same curated segment from the original file, thus enabling a direct A/B comparison between the unprocessed and processed file. 

We construct our crowd-sourced subjective tests in an A/B format and utilize Comparative Mean Opinion Scores (CMOS) \cite{cmos} as the metric of comparison. We utilize Amazon MTurk to crowdsource the results and polled 1300 listeners. 

In Figure 3 (b) and 3 (c), a negative subjective score favors the unprocessed file, a positive subjective score favors the enhanced file, and 0 does not favor either. We show that the lowest bound of $\hat{\rho}$ is either identical or in favor of the enhanced data, while also improving between rounds. Furthermore, the highest bound $\hat{\rho}$ is either identical or in favor of enhanced data while staying approximately equal between rounds. These results indicate to us that the model did not lose performance in transparency and was able to perceptually improve upon “clean speech” after two rounds of curation.

\section{Conclusion}
\label{sec:conclusion}

In this paper, we outlined the Speech Enhancement-based Curation Pipeline (SECP). This pipeline enables any SE model to refine speech data in an iterative fashion to increasing the clean speech corpus size. We also showed how this iterative process produced perceptually favorable speech that improves an SE neural network trained on it.

\newpage

\bibliographystyle{IEEEbib}
\bibliography{main}

\end{document}